# Study of the Thermal Pulsation of AGB Stars

**Ghina M. Halabi**[1]
*American University of Beirut*
*Beirut, Lebanon*
*E-mail:* `gfm01@aub.edu.lb`

A systematic investigation on the third dredge up in a $3M_\odot$, solar metalicity AGB star will be presented. The model evolves from the main sequence up to the Asymptotic Giant Branch (AGB). Intermediate mass stars are important because they contribute significantly via the slow-neutron capture nucleosynthesis. The aim of this work is to gain insight on the behaviour of the AGB star during thermal pulsation. This investigation is based on an extended numerical simulation of the evolutionary phases and full, consistent AGB model calculations. In particular, the convective structure during pulsation will be studied, giving particular emphasis to the analysis of the stability of the Schwarzschild boundary that will eventually determine the occurance of Third Dredge Up (hereafter referred to as TDUP). We provide a brief description of our updated evolutionary code and focus primarily on the obtaining the TDUP after 14 thermal pulses. We elaborate on the non-standard treatment of convection known as "overshooting" and its efficiency in achieving the TDUP event associated with the thermal pulsations (helium shell flashes). However, our knowledge of the physics behind the TDUP event, and the theoretical description of the overshooting mechanism suffer from embarrassing gaps and are, therefore, not properly treated in stellar modelling. This is not surprising, particularly with the oversimplified descriptions of important mixing mechanisms like convection in stellar models. However, since such event drives the $^{13}C$ nucleosynthesis of interest for the following s-process activation, we preset a full assessment of this process and its implications as calculations will further proceed in different directions.

---

[1] Speaker

# 1. Introduction

Several observations detected Carbon stars in the Magellanic Clouds [1], [2], [3], [4] which display enrichment in slow neutron capture elements like Sr, Y, Zr, Ba, La, Ce and Nd [5]. These highly luminous red giants are evolved stars with photospheric ratio C/O > 1. In the Hertzsprung Russell diagram, they are located near the tip of the Asymptotic Giant Branch (AGB) and represent the final stage of evolution of a star that starts the AGB with C/O ≈ 0.5 and progressively modifies its surface composition until it becomes a carbon star with C/O > 1. These stars have an extended hydrogen envelope that is largely unstable against convection. Carbon is synthesized by triple-alpha reactions occurring at the base of the He shell surrounding the core. During the thermal pulsations in AGB stars, the thermonuclear runaway in the helium burning shell induces the formation of a convective zone, which partially overlaps the convective envelope, resulting in the third dredge up (TDUP). The source of neutrons in intermediate mass stars is the $^{13}C(\alpha,n)^{16}O$ reaction, raising the question about the production of $^{13}C$ and posing another challenge in AGB modeling. A hypothesis of extra diffusive mixing between the envelope and the He rich zone was proposed [6], [7], with emphasis on the necessity of extra mixing at all convective boundaries, while others propose extra mixing by an exponentially decaying velocity field at the bottom of the convective envelope only[8]. Other alternatives for this extra mixing whether by gravity waves[9], rotationally induced mixing [10] or thermohaline mixing[11] can also be promising models for this extra mixing mechanism.

The outline of this contribution is the following: we first present briefly in section 2 the used evolutionary code. The thermal pulsations of AGB stars and the TDUP event are discussed in section 3. Conclusions and future work are presented in section 4.

# 2. Stellar evolution code

The calculations presented here are done using a one-dimensional Lagrangian numerical code. This code is described in [12]. The criterion for convection is based on the Schwarzschild criterion $\nabla_{rad} \geq \nabla_{ad}$ where $\nabla_{rad} = \frac{3}{16\pi ac} \frac{\kappa LP}{GmT^4}$. Mixing is performed by a diffusion equation (see [12]). Convective overshooting is included by using the model originally suggested by the hydrodynamical simulations of [13]. These simulations show that convective elements extend beyond the convective boundary determined by the Shwarzschild criterion with an exponentially decaying velocity profile. For the regions immediately below the convective envelope, the diffusion coefficient is parametrized in terms of the pressure scale height $H_p$ as:

$D_{os\ region}(z) = D_{conv\ boundary} e^{\frac{-2z}{fH_p}} \quad z = |r_{boundary} - r|$, where f is a free parameter and z is the radial distance below the boundary determined by the Schwarzschild criterion.

## 3. Evolution up to the thermally pulsating AGB phase

The Asymptotic Giant Branch (referred to as AGB) phase is characterized by the occurrence of thermal pulsations. It is well known that this type of pulsation is caused by the thermal runaway caused by helium burning confined in a relatively thin shell surrounding an inactive carbon-oxygen core. This thermal instability, known as "He-shell flashes", occurs in intermediate mass stars with masses up to $8M_\odot$. In fig. 1, the TP-AGB phase is shown on the HR diagram of a $3M_\odot$ star. These pulsations which are due to an exchange in power between helium shell burning and hydrogen shell burning, are shown in fig. 2. It is also well known that the TDUP occurs during these thermal pulsations of AGB stars.

This process is rather complex from both the physical and numerical points of view. To illustrate this, we discuss in some details the general features characterizing a thermal pulse, with the help of fig. 3.

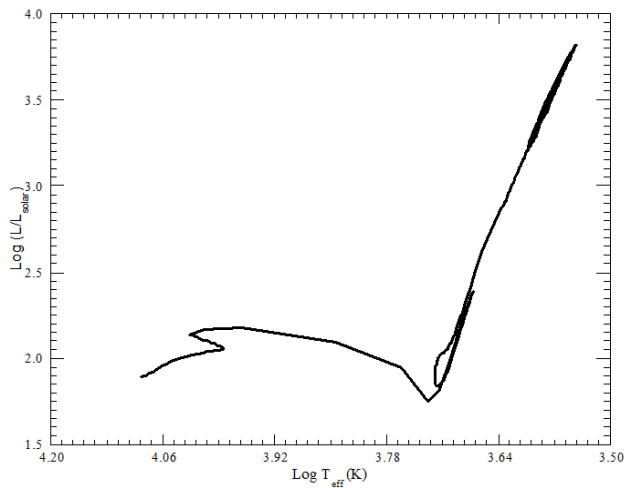
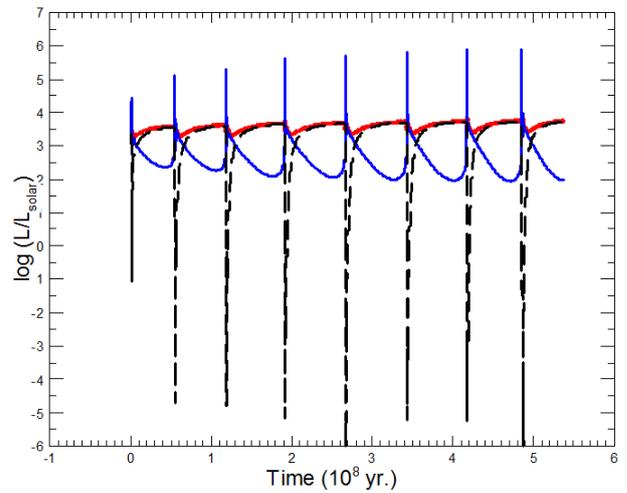

*Figure 1 HR diagram of a $3M_\odot$ star of solar-like initial composition. The luminosity (in solar units) is shown vs. the effective temperature.*

*Figure 2 Stellar luminosity of a $3M_\odot$ star (thick line), helium luminosity (solid line) and hydrogen luminosity (dashed line).*

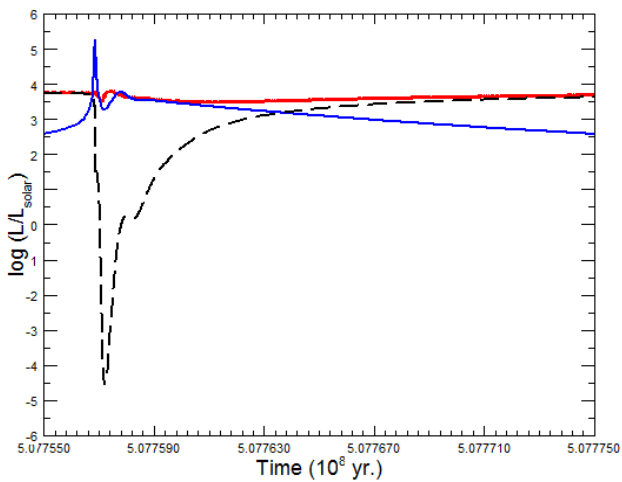

*Figure 3: The helium shell luminosity (solid line with a spike) and the hydrogen shell luminosity (dashed line. Note that the star's luminosity remains essentially constant.*

Fig. 3 shows a detailed view of the one thermal pulse. The four phases of the pulse are: the interpulse phase (off phase) where the quiescent H-burning provides the star's luminosity, it lasts for 5 to $10 \times 10^4$ years. The pulse (on phase), lasts for a few 100 years, where the He-shell flash results in a luminosity peak, and a pulse-driven convective zone (PDCZ) is formed. Then we have the power down phase, when convection shuts off and the H-shell is almost extinguished, then the Third Dredge Up, which corresponds to the point of minimum hydrogen and helium luminosities in fig. 3, where the envelope extends inward into the region previously occupied by the PDCZ.

Fig. 4 shows the convective structure of the 14$^{th}$ and 15$^{th}$ thermal pulses as obtained in the present calculations. It can be seen that the helium flash leads to the formation of a convective zone (PDCZ). Due to the short lifetime of these pulses, high resolution is required to resolve their structure. The time step is decreased to about 100 hours during the pulse. Spacial resolution is also enhanced, where about 2000 mesh points are used for each model during the thermal pulsation. At the pulse, the envelope is driven outward, causing the temperature and density at its bottom to decrease which temporarily turns off the shell hydrogen burning. Shortly after the end of the flash, convection is shut down in the PDCZ, the pushed envelope matter falls back and raises the tempertaure enough to re-ignite the hydrogen shell, causing the convective envelope to retreat as shown in fig. 4. However, the envelope doesn't penetrate deep enough to the region previously occupied by the PDCZ, i.e, the region which is C-rich, to achieve the third dredge up.

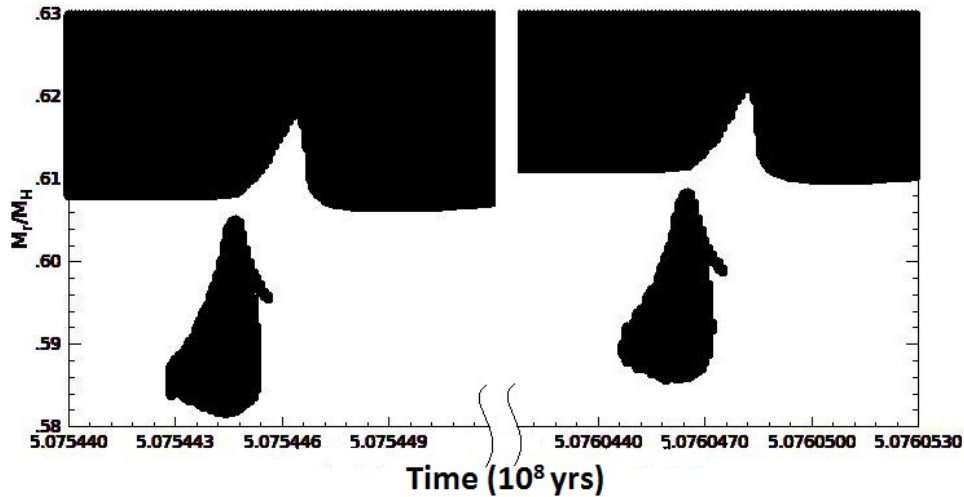

*Figure 4 Two PDCZs associated with the 14$^{th}$ and 15$^{th}$ thermal pulses. Convective regions are shown in black and radiative regions in white. In order to show 2 pulses clearly on the same graph, the time scale between the two pulses is interrupted since the interpulse occurs at a much longer time scale compared to the pulse.*

To assess our case more closely, since we are using the Shwarzschild criterion to determine the convective zones, we compare in fig. 5 the radiative and adiabatic gradients, in addition to the hydrogen and carbon profiles, at the bottom of the convective envelope. It is seen that the convective envelope penetrates down to about 0.6194 $M_\odot$, 200 years after the peak of the 14th pulse (maximum helium luminosity) even when an envelope overshooting of $1H_p$ is used according to the scheme described in section 2.

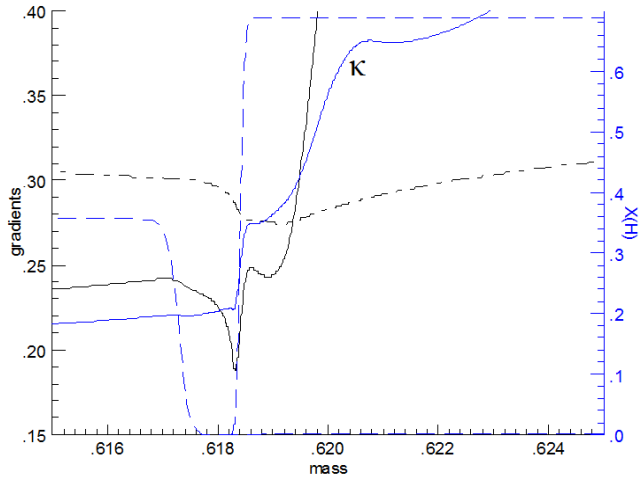

*Figure 5 Radiative gradient (solid line), adiabatic gradient (dot-dashed line), hydrogen and carbon mass fractions (dashed lines) and the opacity profile (labelled κ) at the bottom of the convective envelope when it reaches deepest penetration, 200 years after the peak of the helium flash.*

The reason that the bottom of the convective envelope does not reach the hydrogen discontinuity is due to the drop of $\nabla_{rad}$ below $\nabla_{ad}$ because of the change in the opacity profile which is also shown in fig. 5. This requires careful investigation, and it shows how critical the behavior of the physical quantities is, in a rather narrow mass range, in order to achieve TDUP.
A desirable situation is to have the bottom of the convective envelope at the location of the hydrogen discontinuity, as is the situation with [14]. In this case, the convective envelope will be unstable against convection and even a small amount of extra diffusion of fresh hydrogen would be enough to increase the opacity and thus, invoke a further penetration of the envelope until it reaches the carbon rich zone. While a deeper overshooting seems to be required, at least in the case of solar metalicity, a problem remains to describe it on consistent physical grounds.

It is useful to compare these results with other calculations. Fig. 6 shows the results obtained by [15] for a $2M_\odot$ star with solar-like initial composition. This figure shows that achieving TDUP does not require the bottom of the convective envelope to coincide with the hydrogen discontinuity, unlike the case in [14], it does, however, necessitate envelope overshooting[15].

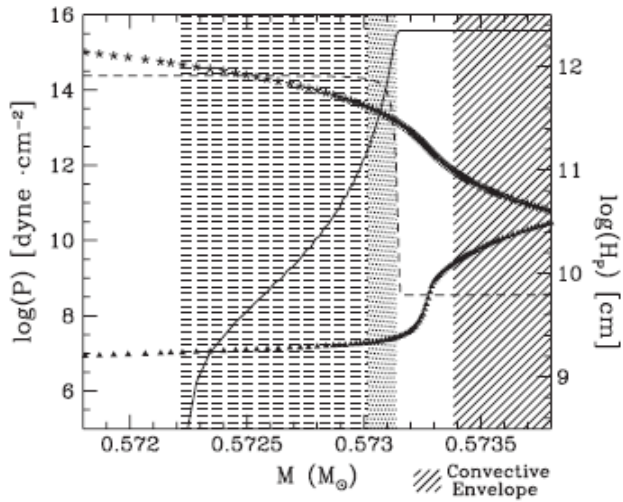

*Figure 6 Core–envelope transition region at the time of maximum penetration for a solar metallicity $2M_\odot$ model. The solid and dashed lines represent the H and the $^{12}C$ profiles, respectively (abundances have been shifted upward in order to match the pressure scale axis). The envelope region is represented by the slant-dashed area. Adopted from [15].*

## 4. Conclusion

Modelling the AGB phase of a star's evolution is challenging from both the physical and computational points of view. The treatment of mixing during this phase still requires further investigation, especially to determine the TDUP. This is an important feature of AGB stars that can not be overlooked because it is responsible for the formation of heavy elements via the s-process mucleosynthesis. Investigation is still carried on in several directions to solve this issue.

## Acknowledgments

This work is part of a Ph-D thesis that I am completing at the American University of Beirut in collaboration with my advisor Dr. Mounib El Eid, for whom I'm grateful for revising this manuscript.